\input{aipcheck}

\documentclass[final]{aipproc}

\newcommand{\beq}{\begin{eqnarray}}
\newcommand{\eeq}{\end{eqnarray}}

\layoutstyle{6x9}

\begin{document}

\title{Extensive nonadditive entropy in quantum spin chains}



\author{Filippo Caruso}{
  address={NEST CNR-INFM \& Scuola Normale Superiore,
  Piazza dei Cavalieri 7, I-56126 Pisa, Italy}
}

\author{Constantino Tsallis}{
  address={Centro Brasileiro de Pesquisas Fisicas, Rua Xavier Sigaud 150,
  22290-180 Rio de Janeiro, Brazil
  \\ and Santa Fe Institute, 1399 Hyde Park Road, Santa Fe, NM 87501,
USA} }

\begin{abstract}
We present details on a physical realization, in a many-body
Hamiltonian system, of the abstract probabilistic structure
recently exhibited by Gell-Mann, Sato and one of us (C.T.), that
the {\it nonadditive} entropy $S_q=k [1- Tr \hat{\rho}^q]/[q-1]$
($\hat{\rho}\equiv$ density matrix; $S_1=-k Tr \hat{\rho} \ln
\hat{\rho}$) can conform, for an anomalous value of $q$
(\textit{i.e.}, $q \neq 1)$, to the classical thermodynamical
requirement for the entropy to be {\it extensive}. Moreover, we
find that the entropic index $q$ provides a tool to characterize
both universal and nonuniversal aspects in quantum phase
transitions (\textit{e.g.}, for a $L$-sized block of the Ising
ferromagnetic chain at its $T=0$ critical transverse field, we
obtain $\lim_{L\to\infty}S_{\sqrt{37}-6}(L)/L=3.56 \pm 0.03$). The
present results suggest a new and powerful approach to measure
entanglement in quantum many-body systems. At the light of these
results, and similar ones for a $d=2$ Bosonic system discussed by
us elsewhere, we conjecture that, for blocks of linear size $L$ of
a large class of Fermionic and Bosonic $d$-dimensional many-body
Hamiltonians with short-range interaction at $T=0$, we have that
the {\it additive} entropy $S_1(L) \propto [L^{d-1}-1]/(d-1)$
({\it i.e.}, $ \ln L$ for $d=1$, and $ L^{d-1}$ for $d>1$), hence
it is {\it not} extensive, whereas, for anomalous values of the
index $q$, we have that the {\it nonadditive} entropy $S_q(L)
\propto L^d$ ($\forall d$), {\it i.e.}, it is extensive. The
present discussion neatly illustrates that {\it entropic
additivity} and {\it entropic extensivity} are quite different
properties, even if they essentially coincide in the presence of
short-range correlations.
\end{abstract}

\maketitle

\section{Introduction}

The appearance of long-range correlations in the ground state of a
quantum many-body system, undergoing a quantum phase transition at
zero temperature, is due to the entanglement \cite{sachdev00}.
Quantum spin chains, composed by a set of localized spins coupled
through short-range exchange interaction in an external transverse
magnetic field, capture the essence of these intriguing phenomena
and have been extensively studied
\cite{osborne02,osterloh02,vidal03,its06,latorre04}. The degree of
entanglement between a block of $L$ contiguous spins and the rest
of the chain in its ground state, as measured by the von Neumann
block entropy $S_1(L) \equiv -k Tr\, \hat{\rho}_L \ln
\hat{\rho}_L$ ($\hat{\rho}_L \equiv Tr_{N-L} \hat{\rho}_N$ is the
reduced density matrix of a $L$-sized block within a $N \to
\infty$ chain with density matrix $\hat{\rho}_N)$, typically
saturates ({\it i.e.}, $\lim_{L \to\infty}S_1(L) < \infty$) or is
logarithmically unbounded ({\it i.e.}, $S_1(L) \propto \ln L$) for
large block size, off or at the critical point, respectively. Here
we show that the {\it nonadditive} entropy
\cite{tsallis88,review05} $S_q(L) \equiv k \frac{1-{\rm Tr}
\hat{\rho}_L^q}{q-1}$ of the block of $L$ spins of the ground
state of quantum spin chains in the neighborhood of a quantum
phase transition is {\it extensive} (\textit{i.e.}, for $L \gg 1$,
$S_q(L)\propto L$) for special values of $q<1$. The {\it additive}
von Neumann entropy $S_1(L)=-k \,{\rm Tr} \hat{\rho}_L \ln
\hat{\rho}_L$ is (like the {\it additive} Renyi entropy) {\it
nonextensive}; indeed, $\lim_{L \to \infty} S_1(L)/L = 0$ in all
considered cases. We present here details of the first physical
realization (this as well as another, Bosonic, physical
realization have been discussed in \cite{CarusoTsallis2007}), in a
many-body Hamiltonian system, of the abstract mathematical
examples shown in Ref. \cite{tsallis05}, that, for anomalous
values of $q$, the {\it nonadditive} entropy $S_q$, can be {\it
extensive}, as expected from the Clausius thermodynamical
requirement for the entropy. We find that the index $q$ provides a
new and efficient tool to characterize different universality
classes in quantum phase transitions, and to quantify entanglement
\cite{nielsen,plenio,horodecki} in quantum many-body systems, by
using a nonadditive measure
\cite{zyczkoski,zeilinger,grigolini,lloyd,abe,abe1,rajagopal,virmani}.

\section{Nonextensive Statistical Mechanics}

The aim of statistical mechanics is to establish a direct link
between the mechanical microscopic laws and classical
thermodynamics. The most famous classical theory in this field has
been developed by Boltzmann and Gibbs (BG) and it is considered
one of the cornerstones of contemporary physics. The connection
between micro- and macro-world is described by the so called BG
entropy:
 \beq
S_{BG} = - k \sum_{i=1}^{W} \,  p_i \ln p_i \label{eq:BGentro}
 \eeq
where $k$ is a positive constant, $W$ is the number of microscopic
states and $\{ p_i \}_{i=1, \ldots , W}$ is a normalized
probability distribution. One of the crucial properties of the
entropy in the context of classical thermodynamics is {\it
extensivity}, namely proportionality with the number of elements
of the system. The BG entropy satisfies this prescription {\it if}
the subsystems are statistically (quasi-) independent, or
typically if the correlations within the system are generically
local. In such cases the system is called \textit{extensive}.

In general, however, the situation is {\it not} of this type and
correlations may be far from negligible at all scales. In such
cases the BG entropy may be nonextensive. Nonetheless, for an
important class of such systems, an entropy exists which is
extensive in terms of the microscopic
probabilities~\cite{tsallis05}. The additive BG entropy can be
generalized into the nonadditive $q$-entropy~\cite{tsallis88} \beq
S_q  = k \frac{1-\sum_{i=1}^W p_i^q}{q-1} \, , \quad q \in {\cal
R}\;\;\;(S_1=S_{BG}) \, . \label{eq:class_ent} \eeq This is the
basis of the so called {\it nonextensive statistical
mechanics}~\cite{review05}, which generalizes the BG theory.

Additivity (for two probabilistically independent subsystems $A$
and $B$) is generalized by the following {\it pseudo-additivity}:
$S_q(A,B)/k=S_q(A)/k+S_q(B)/k+(1-q)S_q(A)S_q(B)/k^2$; the cases $q
< 1$ and $q > 1$ correspond to \textit{super-additivity} and
\textit{sub-additivity}, respectively. For subsystems that have
special probability correlations, extensivity is not valid for
$S_{BG}$, but may occur for $S_q$ with a particular value of the
index $q \ne 1$. Such systems are sometimes referred to as
\textit{nonextensive} \cite{tsallis05,review05}.

A physical system may exhibit genuine quantum aspects. In
particular, quantum correlations, quantified by the entanglement,
can be present. The classical probability concepts are replaced by
the density matrix operator $\hat{\rho}$, in terms of a more
general probability amplitude context. Therefore the quantum
counterpart of the BG entropy in Eq. (\ref{eq:BGentro}), which is
called von Neumann entropy, is given by $S_1 (\hat{\rho}) = - k
{\rm Tr} \, \hat{\rho} \ln \hat{\rho} $, while the classical
$q$-entropy, Eq.~\eqref{eq:class_ent}, is replaced by:
 \beq S_q (\hat{\rho}) =
k \frac{1-{\rm Tr} \, \hat{\rho}^q}{q-1} \, . \label{eq:quant_ent}
 \eeq
The pseudo-additivity property is now given by
 \beq &&\frac{S_q
(\hat{\rho}_1 \otimes \hat{\rho}_2)}{k} = \frac{S_q
(\hat{\rho}_1)}{k} + \frac{S_q (\hat{\rho}_2)}{k} + (1-q)
\frac{S_q (\hat{\rho}_1)}{k} \frac{S_q (\hat{\rho}_2)}{k}\nonumber \;;
\label{eq:pseudoadd}
 \eeq
from now on $k=1$.

\section{XY Model}

In this paper we analyze a quantum system in which strong
non--classical correlations occur between its components. We focus
our investigations on a one-dimensional spin-$1/2$ ferromagnetic
chain with an exchange (local) coupling and in the presence of an
external transverse magnetic field, \textit{i.e.}, the quantum XY
model. The Hamiltonian of the XY model with open boundary
conditions is:
 \beq \hat{\mathcal H} =
- \sum_{j=1}^{N-1} \left[ (1 + \gamma) \hat{\sigma}^x_j
\hat{\sigma}^x_{j+1} + (1 - \gamma) \hat{\sigma}^y_j
\hat{\sigma}^y_{j+1} + 2 \lambda \hat{\sigma}^z_j \right]
\label{eq:XYham} \eeq where $\hat{\sigma}^\alpha_j (\alpha =
x,y,z)$ are the Pauli matrices of the $j$-th spin, $N$ is the
number of spins of the chain, $\gamma$ and $\lambda$ characterize,
respectively, the strength of the anisotropy parameter and of a
trasverse magnetic field along the $z$ direction. This model for
$0 < \vert \gamma \vert \leq 1$ belongs to the Ising universality
class and it actually reduces to the quantum Ising chain for
$|\gamma|=1$. This system undergoes a quantum phase transition at
the critical point $|\lambda_c| = 1$ in the thermodynamic limit $N
\to \infty$. For $\gamma=0$ it is the isotropic XX model, which is
critical for $\vert \lambda \vert \leq 1$~\cite{sachdev00}. Let us
stress that in the following we will solve analytically the ground
state in the limiting case of an infinite chain, i.e. $N \to
\infty$. Therefore, the coupling being ferromagnetic, the results
will not depend on our particular choice of the boundary
conditions.

The entanglement in the neighborhood of the quantum phase
transition has been recently widely investigated, thus
establishing a direct connection between quantum information
theory and condensed matter
physics~\cite{osterloh02,osborne02,vidal03,its06,latorre04}. In
particular it has been shown that one-site and two-site
entanglement between nearest or next-to-nearest spins display a
peak near or at the critical point~\cite{osterloh02,osborne02}. On
the other side, the entanglement between a block of $L$ contiguous
spins and the rest of the chain in the ground state, quantified by
the von Neumann entropy, presents a logarithmic divergence with
$L$ at criticality, while it saturates in a non--critical
regime~\cite{vidal03,its06,latorre04}.

The inadequacy of the additive von Neumann entropy as a measure of
the information content in a quantum state has been pointed out in
Ref. \cite{zeilinger}. A theoretical observation that the measure
of quantum entanglement may not be additive has been discussed in
Refs.
\cite{grigolini,zyczkoski,zeilinger,lloyd,abe,abe1,rajagopal,virmani}.
Recently, Ref. \cite{popescu} suggested to abandon the \textit{a
priori} probability postulate going beyond the usual BG situation.

Here we propose to extend the definition of the von Neumann
entropy to a wider class of entropy measures which naturally
include it, thus generalizing the notion of the block entanglement
entropy. The block $q$-entropy of a block of size $L$ is simply
defined as the $q$-entropy, Eq.~\eqref{eq:quant_ent}, of the
reduced density matrix $\hat{\rho}_L$ of the block, when the total
chain is in the ground state. In the following we show that,
contrary to the von Neumann entropy, there exists a $q$ value for
which $S_q(\hat{\rho}_L)$ is {\it extensive}. This value does
depend on the critical properties of the chain and it is
consistent with the universality hypothesis.

The XY model in Eq.~\eqref{eq:XYham} can be diagonalized exactly
with a Jordan-Wigner transformation, followed by a Bogoliubov
rotation~\cite{lieb61,pfeuty,barouch,barouch1}; this allows one to
analytically consider the thermodynamic limit $N \to \infty$. The
normal modes of the system are linear combinations of the
following non--local Majorana fermions:
 \beq \hat{c}_{2l} \equiv
\left( \prod_{k=0}^{l-1} \hat{\sigma^z_l} \right) \hat{\sigma}^x_l
\, ; \qquad \hat{c}_{2l+1} \equiv \left( \prod_{k=0}^{l-1}
\hat{\sigma^z_l} \right) \hat{\sigma}^y_l \, . \label{eq:majorana}
 \eeq
These operators are Hermitian and obey the anti-commutation rules
$\{ c_m, c_n \} = 2 \delta_{m n}$. The ground state $\vert \Psi_g
\rangle$ is completely characterized by the scalar product
$\langle c_m c_n \rangle \equiv \delta_{mn} + i \,
\Gamma^{(N)}_{mn}$, where
 \beq \Gamma^{(N)} = \left[
\begin{array}{cccc}
\Pi_0     & \Pi_1  & \cdots & \Pi_{N-1} \\
\Pi_{-1}  & \Pi_0  &   {}   & \vdots    \\
\vdots    &   {}   & \ddots & \vdots    \\
\Pi_{1-N} & \cdots & \cdots & \Pi_0
\end{array} \right] , \,
\Pi_l = \left[ \begin{array}{cc}
0       &  g_l \\
-g_{-l} &  0
\end{array} \right] \nonumber
 \eeq
with real coefficients $g_l$ given, for an infinite chain, by $
g_l = \frac{1}{2 \pi} \int_{0}^{2 \pi} d \phi e^{-i l \phi}
\frac{\cos \phi - \lambda - i \gamma \sin \phi} {\vert \cos \phi -
\lambda - i \gamma \sin \phi \vert}$. The spectrum of
$\hat{\rho}_L$ in an infinite chain in its ground state can then
be exactly evaluated~\cite{latorre04}. Indeed, the matrix
$\hat{\rho}_L$ can be written as a tensor product in terms of $L$
uncorrelated non--local Fermionic modes, which are linear
combinations of the operators $\hat{c}_n$ in
Eq.~\eqref{eq:majorana}: $\hat{\rho}_L = \hat{\tau}_1 \otimes
\ldots \otimes \hat{\tau}_L$, where $\hat{\tau}_l$ denotes the
mixed state of mode $l$. The eigenvalues of $\hat{\tau}_l$ are $(1
\pm \nu_l )/2$, where $\nu_l$ is the imaginary part of the
eigenvalues of the matrix $\Gamma^{(L)}$. The entropy in
Eq.~\eqref{eq:quant_ent} is then easily computed by using the
pseudo-additivity, Eq.~\eqref{eq:pseudoadd}, and by noticing that
the trace of $\hat{\tau}_l^q$ is simply ${\rm Tr \ \hat{\tau}_l^q}
= [(1 + \nu_l)/2]^q + [(1 - \nu_l)/2]^q$. Notice that the required
computational time scales polynomially with the block size $L$,
thus allowing one to reliably analyze blocks with up to a few
hundreds of spins.

\section{Results}

\begin{figure}[h!]
\includegraphics[width=0.6\textwidth]{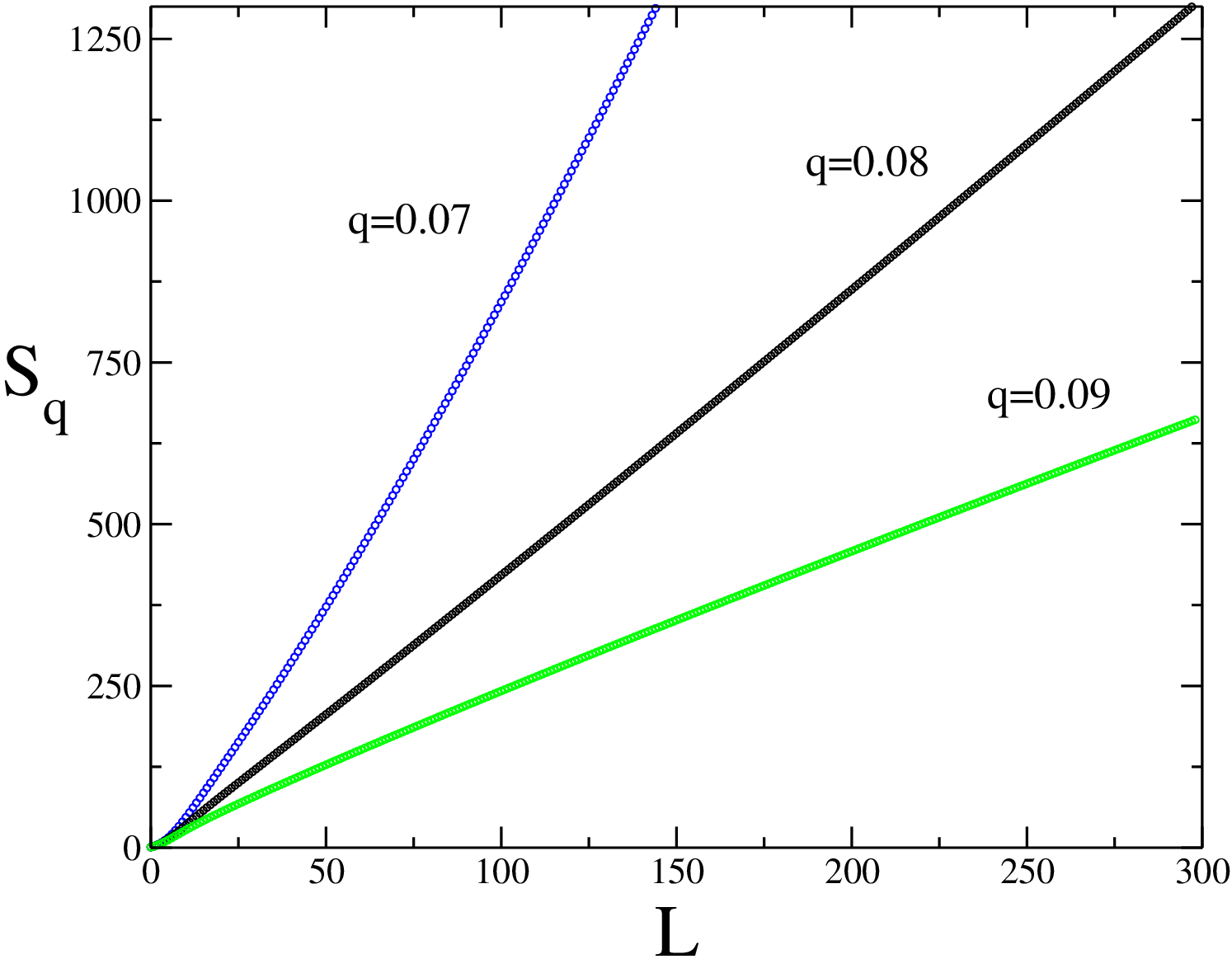}
\caption{Block $q$-entropy $S_q(\hat{\rho}_L)$ as a function of
the block size $L$ in a critical Ising chain ($\gamma=1, \,
\lambda=1$), for typical values of $q$. Only for $q=q_{ent} \simeq
0.0828$, $s_q$ is {\it finite} (\textit{i.e.}, $S_q$ is {\it
extensive}); for $q < q_{ent}$ ($q > q_{ent}$) it diverges
(vanishes).}\label{fig.1}
\end{figure}

We first analyze the anisotropic quantum XY model,
Eq.~\eqref{eq:XYham} with $\gamma \neq 0$, that has a critical
point in $\lambda_c = 1$. The block $q$-entropy as a function of
the block size can show completely different asymptotic behaviors,
by varying the entropic index $q$. In particular, here we are
interested in a thermodynamically relevant quantity, namely the
slope, noted $s_q$, of $S_q$ versus $L$. It is generically not
possible to have a finite value of $s_1$: the entanglement
entropy, evaluated by the von Neumann entropy, either saturates or
diverges logarithmically in the thermodynamic limit, for
respectively non--critical or critical spin chains
\cite{vidal03,its06,latorre04}. The situation dramatically changes
by using the entropy in Eq.~\eqref{eq:quant_ent}: qualitatively it
happens that, regardless the presence or absence of criticality, a
$\lambda$-dependent value of $q$, noted $q_{ent}$, exists such
that, in the range $1 \ll L \ll \xi$ ($\xi$ being the correlation
length), $s_{q_{ent}}$ is finite, whereas it vanishes (diverges)
for $q>q_{ent}$ ($q<q_{ent}$). We note that here the
\textit{nonextensivity} (\textit{i.e.}, $q\neq 1$) features are
not due to the presence of say long-range interactions
\cite{anteneodo} but they are triggered only by the fully quantum
{\it nonlocal} correlations. In Fig. \ref{fig.1} we show, for the
critical Ising model ($\lambda=1$, $\gamma=1$), the behavior of
the block $q$-entropy with respect to the block size:
$S_q(\hat{\rho}_L)$ becomes {\it extensive} (\textit{i.e.}, $0<
\lim_{L \to \infty} S_q(\hat \rho_L)/L < \infty$) for $q_{ent}
\simeq 0.0828 \pm 10^{-4}$ (with a corresponding entropic density
$s_{q_{ent}} \approx 3.56 \pm 0.03$), thus satisfying the
classical thermodynamical prescription.

\begin{figure}[h!]
\includegraphics[width=0.6\textwidth]{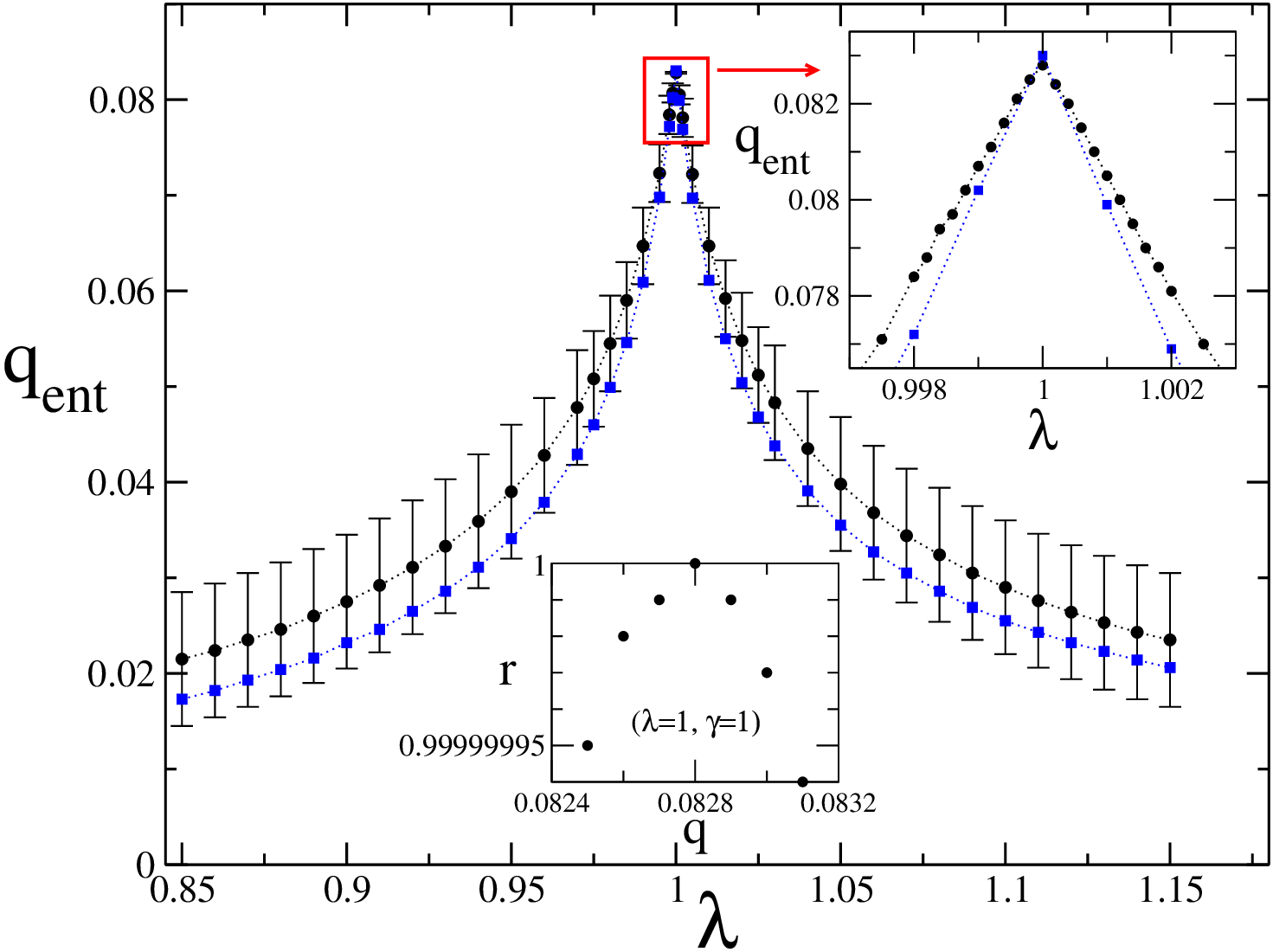} \caption{The
$\lambda$-dependence of the index $q_{ent}$ in the Ising
($\gamma=1$, circle) and XY ($\gamma=0.75$, square) chains. {\it
At bottom}: Determination of $q_{ent}$ through numerical
maximization of the linear correlation coefficient $r$ of
$S_q(\hat{\rho}_L)$. The error bars for the Ising chain are
obtained considering the variation of $q_{ent}$ when using the
range $100 \le L \le 400$ in the search of $S_q(\hat{\rho}_L)$
linear behavior. Actually, at the present numerical level, we
cannot exclude finite-size effects off criticality.}\label{fig.2}
\end{figure}

A very similar behavior is shown for non--critical Ising model, as
well as for critical and non--critical XY models with
$0<\gamma<1$. The value of $q_{ent}$, for which
$S_q(\hat{\rho}_L)$ is {\it extensive}, is obtained maximizing
numerically the linear correlation coefficient $r$ of
$S_q(\hat{\rho}_L)$, in the range $1 \ll L \ll \xi$, with respect
to $q$, as shown in the bottom inset in Fig. \ref{fig.2}. Let us
stress that, at precisely the critical point, $\xi$ diverges,
hence $L$ is unrestricted and can run up to infinity. The index
$q_{ent}$ depends on the distance from criticality and it
increases as $\lambda$ approaches $\lambda_c$ (Fig. \ref{fig.2}).
It is worth stressing that our numerical results satisfy the
duality symmetry $\lambda \longrightarrow 1/\lambda$, investigated
in Ref. \cite{savit}.

\begin{figure}[h!]
\includegraphics[width=0.6\textwidth]{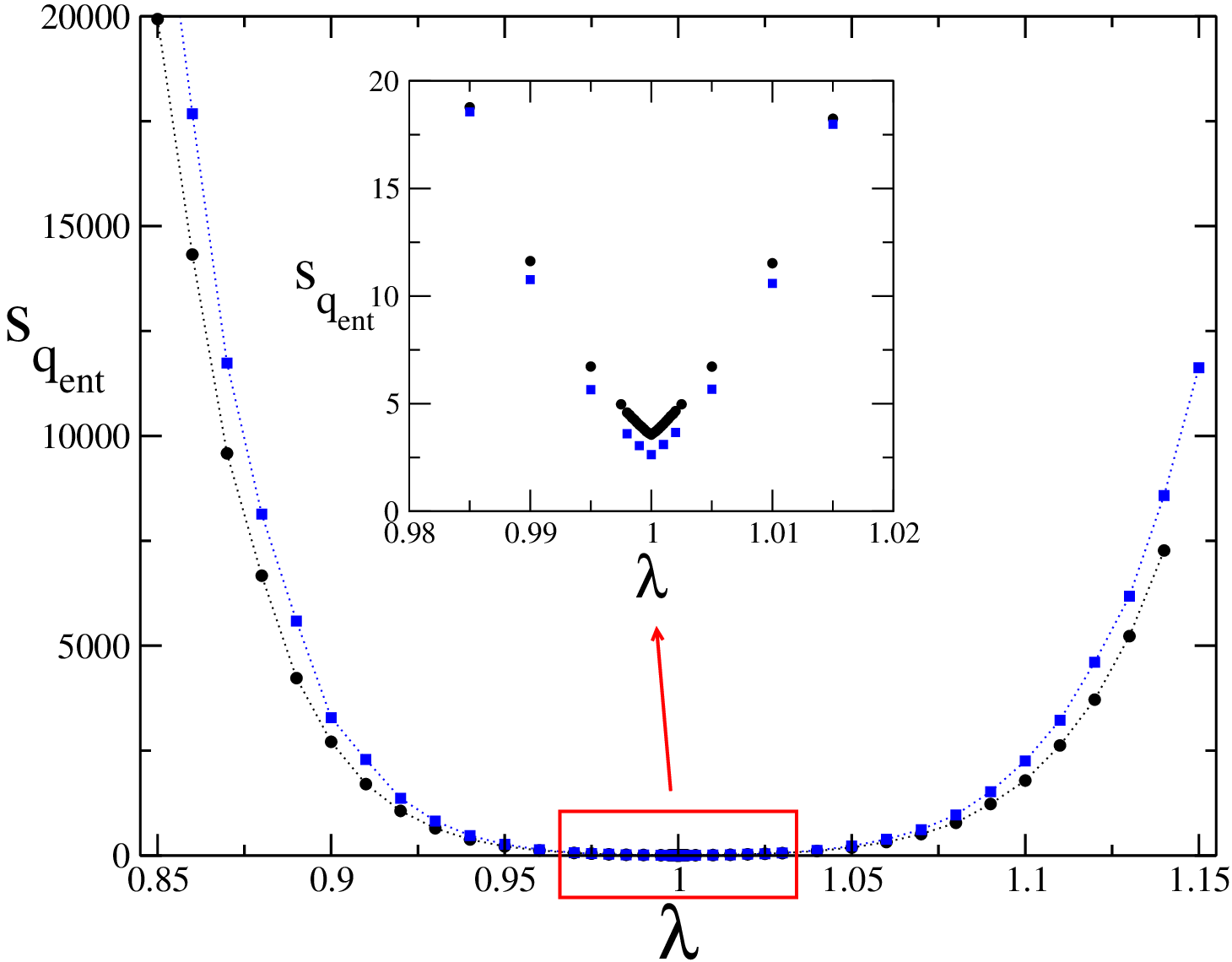} \caption{The
$\lambda$-dependence of the $q$-entropic density $s_{q_{ent}}$ in
the Ising ($\gamma = 1$, circle) and XY ($\gamma = 0.75$, square)
models. For $\lambda=1$, the slopes are 3.56 and 2.63, for $\gamma
=1$ and $\gamma=0.75$, respectively. }\label{fig.3}
\end{figure}

We have also checked other values of $\gamma$ for the XY model and
the results are very similar to those presented here. This fact is
consistent with the universality hypothesis. On one hand, XY and
Ising model (Ising universality class) have the same behavior as
regards the extensivity of $S_q(\hat{\rho}_L)$; in Fig.
\ref{fig.3} we report the variation of $s_{q_{ent}}$ with respect
to $\lambda$. On the other hand, for the isotropic XX model
($\gamma=0$) in the critical region $\vert \lambda \vert \leq 1$
we find $q_{ent}\simeq 0.15 \pm 0.01$ ($\simeq 2 q_{ent}^{XY}$
with $q_{ent}^{XY} \simeq 0.08$) for which $S_{q}(\hat \rho_L)$
becomes {\it extensive}.

\begin{figure}[b!]
\includegraphics[width=0.6\textwidth]{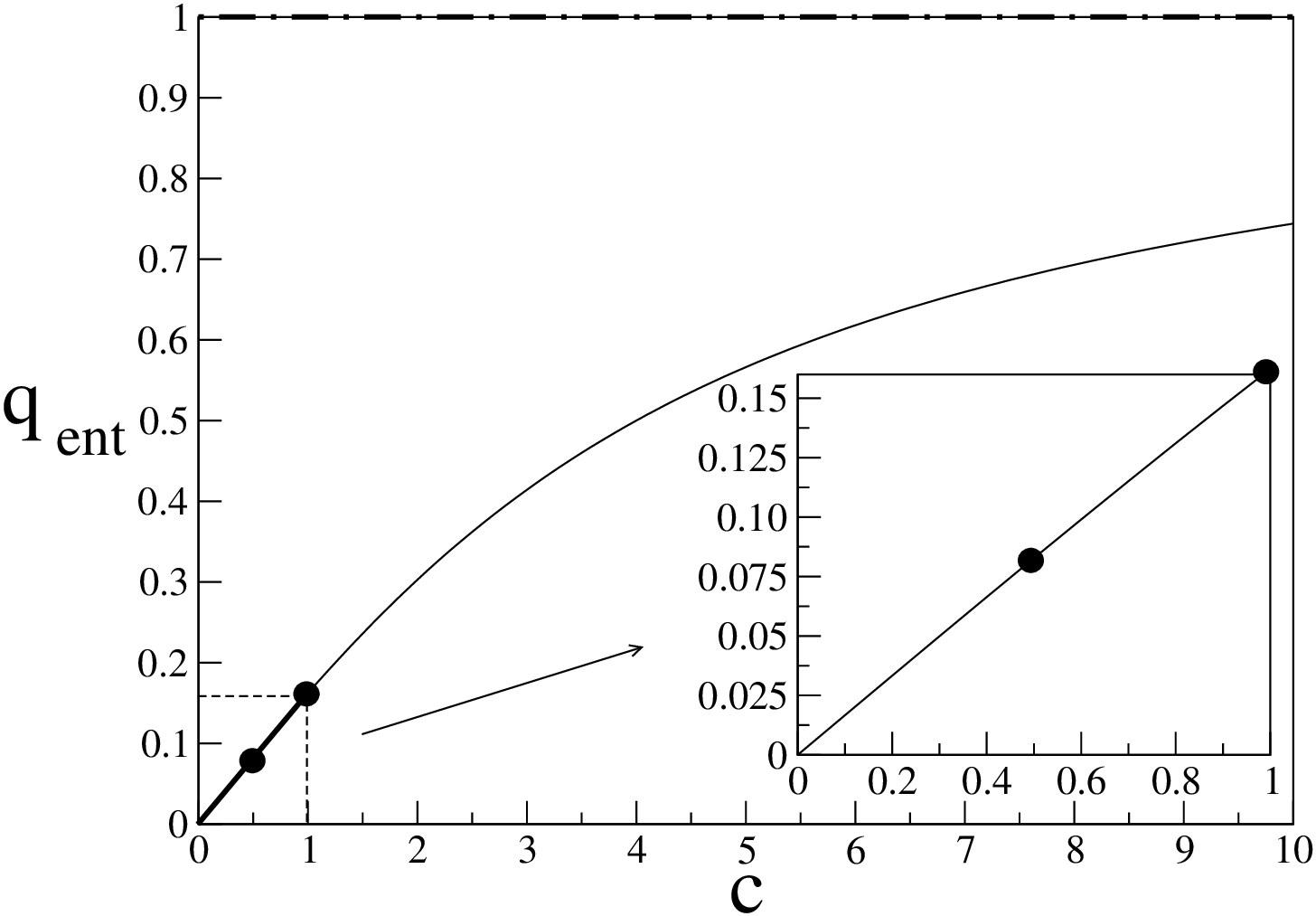} \caption{$q_{ent}$ versus
$c$ with the $q$-entropy, $S_q(\hat{\rho}_L)$, being
\textit{extensive}, \textit{i.e.}, $\lim_{L
\to\infty}S_{\frac{\sqrt{9+c^2}-3}{c}}(\hat{\rho}_L)/L< \infty$.
When $c$ increases from $0$ to infinity, $q_{ent}$ increases from
$0$ to unity (von Neumann entropy); for $c=4$, $q=1/2$ and for $c
\gg 1$, see Ref. \cite{ginsparg}. {\it Inset}: for the critical
quantum Ising and XY models $c=1/2$ and $q_{ent} = \sqrt{37} -6
\simeq 0.0828$, while  for the critical isotropic XX model $c=1$
and $q_{ent} = \sqrt{10} -3 \simeq 0.16$. }\label{fig.4}
\end{figure}

Ref. \cite{cardy} enables us to analytically confirm, at the
critical point, our numerical results. The continuum limit of a
(1+1)-dimensional critical system is a conformal field theory with
central charge $c$. In this quite different context, the authors
re-derive the result $S_1(\hat{\rho}_L)\sim (c/3) \ln L$ for a
finite block of length $L$ in an infinite critical system. To
obtain the von Neumann entropy, they find an analytical expression
for ${\rm Tr} \hat{\rho}^q_L$, namely ${\rm Tr} \hat{\rho}_L^q
\sim L^{-c/6(q-1/q)}$. Here, we use this expression quite
differently. We impose the \textit{extensivity} of
$S_q(\hat{\rho}_L)$ finding the value of $q$ for which
$-c/6(q_{ent}-1/q_{ent})=1$, \textit{i.e.},
\begin{equation}
q_{ent} = \frac{\sqrt{9+c^2}-3}{c}. \label{charge}
\end{equation}
Consequently, $\lim_{L
\to\infty}S_{\frac{\sqrt{9+c^2}-3}{c}}(L)/L< \infty$. When $c$
increases from $0$ to infinity (see Fig. \ref{fig.4}), $q_{ent}$
increases from $0$ to unity (von Neumann entropy); for $c=4$
(dimension of physical space-time), $q=1/2$; $c =26$ corresponds
to a $26$-dimensional Bosonic string theory, see \cite{ginsparg}.
It is well known that for critical quantum Ising and XY models the
central charge is equal to $c=1/2$ (indeed they are in the same
universality class and can be mapped to a free Fermionic field
theory). For these models, at $\lambda=1$, the value of $q$ for
which $S_q(\hat{\rho}_L)$ is {\it extensive} is given by $q_{ent}
= \sqrt{37} -6 \simeq 0.0828$, in perfect agreement with our
numerical results in Fig. \ref{fig.2}. The critical isotropic XX
model ($\gamma=0$ and $|\lambda| \leq 1$) is, instead, in another
universality class, the central charge is $c=1$ (free Bosonic
field theory) and $S_q(\hat{\rho}_L)$ is {\it extensive} for
$q_{ent} = \sqrt{10} -3 \simeq 0.16$, as found also numerically.
We finally notice that, in the $c \to\infty$ limit, $q_{ent} \to
1$. We do not clearly understand the physical interpretation of
this fact. However, since $c$ in some sense plays the role of a
dimension (see \cite{ginsparg}), this limit could correspond to
some sort of mean field approximation. If so, it is along a line
such as this one that a mathematical justification could emerge
for the widely spread use of BG concepts in the discussion of
mean-field theories of spin-glasses (within the replica-trick and
related approaches). Indeed, BG statistical mechanics is
essentially based on the {\it ergodic} hypothesis. It is firmly
known that glassy systems (e.g., spin-glasses) precisely {\it
violate ergodicity}, thus leading to an intriguing and fundamental
question. Consequently, a mathematical justification for the use
of BG entropy and energy distribution for such complex mean-field
systems would be more than welcome.

It is worth to mention that the Renyi entropy of a block of
critical XX spin chains has been derived analytically in Ref.
\cite{jin04}. Since the Renyi entropy is simply connected to the
entropy $S_q$, it is possible to re-derive $q_{ent}$ for the
critical XX model also from that analytical expression.

\section{Concluding remarks}

Summarizing, we have presented: (i) Details on the first physical
realization (in a $1/2$-spin $d=1$ quantum system), in a many-body
Hamiltonian system, of the abstract probabilistic structure shown
in Ref. \cite{tsallis05}, that $S_q$ conforms, for a special value
of $q$, to the classical thermodynamical requirement for the
entropy to be extensive (the second physical realization, in a
$d=2$  Bosonic system, can be seen in \cite{CarusoTsallis2007});
(ii) A new connection, Eq. (\ref{charge}), between nonextensive
statistical mechanical concepts and BG statistical mechanics at
criticality (see \cite{robledo} for another such analytical
connection); (iii) A novel and simple manner to characterize
entanglement through the pair $(q_{ent},s_{q_{ent}})$.

Let us point out also that the reduction of the {\it pure}
ground state of the full chain (at $T=0$) to a finite block of $L$
spins results in a {\it  mixed} state with quantum fluctuations. A
mapping of this subsystem within a zero temperature XX infinite
chain to a finite system which is thermalized at some finite
temperature has been recently exhibited \cite{eisler}, thus
defining an $L$-dependent effective temperature of the block. The
use of a non-Boltzmannian distribution (\textit{e.g.}, the one
emerging within nonextensive statistical mechanics) might enable
defining an effective temperature which would {\it not} depend on
$L$, as physically desirable. Indeed, this approach has been
successfully implemented for $e-e^+$ collision experiments
\cite{curado}.

Finally, let us emphasize the difference between {\it additivity}
and {\it extensivity} for the entropy. Additivity only depends on
the mathematical definition of the entropy; therefore, $S_1$ is
additive, whereas $S_q$ ($q \ne 1$) is nonadditive. Extensivity is
more subtle, since it also depends on the specific system. The
$T=0$ block entropies of the present $1/2$-spin $d=1$ quantum
system at criticality are given by $S_1(L) \propto \ln L$ ({\it
i.e.}, nonextensive), and $S_{[\sqrt{9+c^2}-3]/c}(L) \propto L$
({\it i.e.}, extensive). It is known (see \cite{Barthel} and
references therein) that, for $d$-dimensional Bosonic systems
(e.g., a black hole \cite{Srednicki}), $S_1$ follows the {\it area
law}, {\it i.e.}, $S_1(L) \propto L^{d-1}$  ({\it i.e.},
nonextensive). A logarithmic behavior for $d=1$, and the area law
for $d>1$ can be unified through $S_1(L) \propto [L^{d-1}-1]/(d-1)
\equiv \ln_{2-d} L$ \cite{qlog} ({\it i.e.}, nonextensive), which
would correspond to a large class (yet not completely identified)
of fully entangled quantum systems. For all these systems, one
could expect that a value of $q$ exists such that $S_q(L) \propto
L^d$ ({\it i.e.}, extensive). In addition to the present example,
a $d=2$ Bosonic system has been shown \cite{CarusoTsallis2007} to
satisfy this conjecture.

\begin{theacknowledgments}
The present work has benefited from fruitful comments by R. Fazio,
V. Giovannetti, D. Patan\`{e}, A. Pluchino, A. Rapisarda and D.
Rossini. One of us (C.T.) also acknowledges enlightening
conversation with G. 't Hooft on the black-hole area law. This
work was partially supported by Centro di Ricerca Matematica E. De
Giorgi of the Scuola Normale Superiore, and by Pronex/MCT, CNPq
and Faperj (Brazilian Agencies). In the 100th anniversary of his
birthday, we dedicate this work to the memory of Ettore Majorana.
\end{theacknowledgments}

\end{document}